\documentclass{article}

\usepackage{amsmath}
\usepackage{natbib}
\usepackage{amssymb}
\usepackage{array}
\usepackage{amsbsy}
\usepackage{amsfonts}
\usepackage{dsfont}
\usepackage{graphicx}
\newtheorem{lem}{Lemma}
\newtheorem{thm}{Theorem}
\newtheorem{defn}{Definition}
\newcommand{\bm}[1]{{\boldsymbol {#1}}}
\begin{document}

\title{Functional Uniform Priors for Nonlinear Modelling}

\author{Bj\"{o}rn Bornkamp\footnote{bjoern.bornkamp@novartis.com}}


\maketitle

\begin{abstract}

  This paper considers the topic of finding prior distributions when a
  major component of the statistical model depends on a nonlinear
  function.  Using results on how to construct uniform distributions
  in general metric spaces, we propose a prior distribution that is
  uniform in the space of functional shapes of the underlying
  nonlinear function and then back-transform to obtain a prior
  distribution for the original model parameters. The primary
  application considered in this article is nonlinear regression, but
  the idea might be of interest beyond this case. For nonlinear
  regression the so constructed priors have the advantage that they
  are parametrization invariant and do not violate the likelihood
  principle, as opposed to uniform distributions on the parameters or
  the Jeffrey's prior, respectively.  The utility of the proposed
  priors is demonstrated in the context of nonlinear regression
  modelling in clinical dose-finding trials, through a real data
  example and simulation. In addition the proposed priors are used for
  calculation of an optimal Bayesian design.
\end{abstract}

\section{Introduction}
\label{sec:introduction}

Mathematical models of the real world are typically nonlinear,
examples in medical or biological applications can be found for
instance in \citet{lind:2001} or \citet{jone:plan:slee:2010}. Setting
up prior distributions in a statistical analysis of nonlinear models,
however often remains a challenge. If external, numerical or
non-numerical information exists, one can try to quantify it into a
probability distribution, see for example the works of O'Hagan et
al. (2006), \nocite{ohag:2006}\citet{born:icks:2009b}, and
\citet{neue:capk:bran:2010}. The classical approach in the absence of
substantive information is Jeffreys prior distribution (or variants),
given by $p\left(\bm \theta \right) \propto \sqrt{{\det(I(\bm
    \theta))}}$, where $\bm \theta \in \bm \Theta \subset
\mathbb{R}^p$ is the parameter, and $I\left(\bm \theta \right)$ the
Fisher information matrix of the underlying statistical model. See
\citet{kass:wass:1996}, \citet[ch. 5]{ghos:dela:sama:2006} or
\citet{berg:bern:sun:2009} for this approach and generalizations.  A
serious drawback is the fact that this prior can depend on observed
covariates. In the case of nonlinear regression analysis, the prior
depends on the design points and relative allocations to these points
and thus violates the likelihood principle. Apart from the
foundational issues this raises (see, \textit{e.g.},
\citet[ch. 3]{ohag:fors:2004}) it also has undesirable practical
consequences. For Bayesian optimal design calculations in nonlinear
regression models, for example, Jeffreys prior cannot be used, because
it depends on the design points, which is what we want to calculate in
the optimal design problem. In the context of adaptive dose-finding
clinical trials, patients are allocated dynamically to the doses
available (see the works of \citet{muel:berr:grie:2006} or Dragalin et
al. (2010)\nocite{drag:2010}) so that the sequential analysis of the
data will differ from the analysis combining all data, when using
Jeffreys rule. In summary the main issue with the Jeffreys prior
distribution is that one cannot state it before data collection, which
is crucial in some applications. Surprisingly few proposals have been
made to overcome this situation. In current practice often uniform
distributions for $\bm \theta$ on a reasonable compact subset of the
parameter space are used. This approach is however extremely sensitive
to the chosen parametrization (which might be more or less arbitrary)
and can be much more informative than one would expect intuitively.

To illustrate the point, we will use a simple example. Suppose one
would like to analyse data using the exponential model $\exp(-\theta
x)$, here with $x\in [0,10]$, which could be the mean function in a
regression analysis. Assume that no historical data or practical
experiences related to the problem are available.

\begin{figure}
  \centerline{%
  \includegraphics[width=0.9\textwidth]{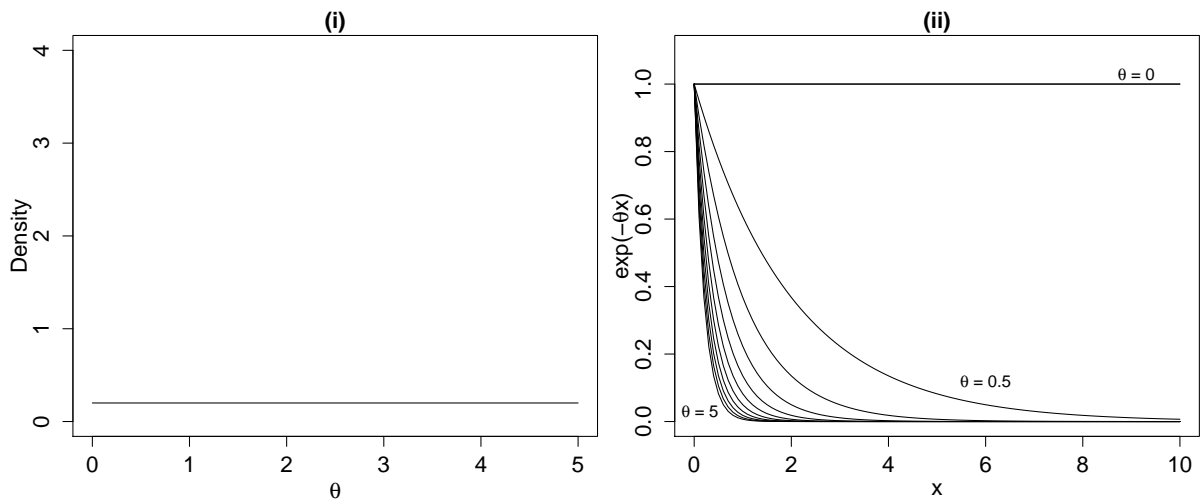}}
 \caption{(i) Display of the uniform distribution on $\theta$ scale;
   (ii) Display of the regression function $\exp(-\theta x)$
   for $\theta=0$, $\theta = 5$ and the $\theta$ corresponding to the
   $i/10$ quantile $i=1,\ldots,9$ of the uniform distribution.}
 \label{fig:expo1}
\end{figure}

A first pragmatic approach in this situation is to use a uniform
distribution on $\theta$ values leading to a reasonable shape coverage
of the underlying regression function $\exp(-\theta x)$, for example
the interval $\theta \in [0,5]$ covers the underlying shapes almost
entirely. The consequences of assuming a uniform prior on $[0,5]$ can
be observed in Figure \ref{fig:expo1} (ii). While the prior is uniform
in $\theta$ space, it places most of its prior probability mass on the
functional shapes that decrease quickly towards zero, and we end up
with a very informative prior distribution in the space of functional
shapes. This is highly undesirable when limited prior information
regarding the shape is available. In addition it depends crucially on
the upper bound selected for $\theta$, and a uniform distribution in
an alternative parameterization would lead to entirely different prior
in the space of shapes. One way to overcome these problems is to use a
distribution that is uniform in the space of functional shapes of the
underlying nonlinear function. This will be uninformative from the
functional viewpoint and will not depend on the selected
parameterization.

In finite dimensional situations it is a standard approach to use
distributions that are uniform in an interpretable parameter
transformation, when it is difficult to use the classical default
prior distributions. In the context of Dirichlet process mixture
modelling, one can use a uniform distribution on the probability that
two observations cluster into one group and then transfer this into a
prior distribution for the precision parameter of the Dirichlet
process. In the challenging problem of assigning a prior distribution
for variance parameters in hierarchical models, \citet{dani:1999}
assumes a uniform distribution on the shrinkage coefficient and then
transfers this to a prior distribution for the variance parameter. In
these cases the standard change of variables theorem can be used to
derive the necessary uniform distributions. When we want to impose a
uniform distribution in the space of functional shapes of an
underlying regression function, however, it is not entirely obvious
how to construct a uniform distribution. In the next section we will
review a methodology that allows to construct uniform distributions on
general metric spaces.  In Section \ref{sec:nonreg} we will adapt this
to the nonlinear models that we consider in this article. Finally in
Section \ref{sec:appl} we test the priors for nonlinear regression on
a data set from a dose-finding trial, a simulation study and an
optimal design problem.

\section{Methodology}
\label{sec:methodology}

\subsection{General Approach}
\label{sec:general-approach}

Suppose one would like to find a prior distribution for a parameter
$\bm  \theta$ in a compact subspace $\bm  \Theta \subset
\mathbb{R}^p,\,p < \infty$. The approach proposed in this paper is to
map the parameter $\bm  \theta$ from $\bm  \Theta$ into another compact
metric space $(M,d)$, with metric $d$, using a differentiable
bijective function $\varphi: \bm  \Theta \mapsto M$, 
so that $\varphi(\bm \theta) = \bm \phi \in M$. The metric $d$ should
ideally define a reasonable measure of closeness and distance between
the parameters, and its choice will of course be model and application
dependent. In the exponential regression example, for instance, it
seems adequate to measure the distance between two parameter values
$\theta'$ and $\theta''$ by a distance between the resulting functions
$\exp(-x\theta')$ and $\exp(-x\theta'')$, rather than the Euclidean
distance between the plain parameter values. In this metric space
$(M,d)$, one then imposes a uniform distribution, reflecting the
appropriate notion of distance of the metric space $(M,d)$, and
transforms this distribution back to the parameter scale.

The construction of a uniform distribution in general metric spaces
has been described by \citet{demb:1990}, using the notion of packing
numbers. \citet{ghos:ghos:rama:1997} apply this result for two
particular Bayesian applications (derivation of Jeffreys prior for
parametric problems and nonparametric density estimation). In the
following we review and adapt this theory to our situation. Some basic
mathematical notions are needed to present the ideas: Define an
$\epsilon$-net as a set $S_\epsilon \subset M$, so that for all $\bm
\phi',\bm \phi'' \in S_\epsilon$ holds $d(\bm \phi',\bm \phi'') \geq
\epsilon$, and the addition of any point to $S_\epsilon$ destroys this
property. An $\epsilon$-lattice $S_\epsilon^m$ is the $\epsilon$-net
with maximum possible cardinality. Dembski defines the uniform
distribution on $M$ as the limit of a discrete uniform distribution on
an $\epsilon$-lattice on $M$, when $\epsilon \rightarrow 0$.

\begin{defn}
  \label{eqn:def}
  The uniform distribution $\Pi$ on $M$ defined as
  $$\Pi(A)=\underset{\epsilon \rightarrow 0}{\lim}\,\Pi_\epsilon(A),$$
  for $A\subset M$ and $\Pi_\epsilon(A)$ is the discrete uniform
  distribution supported on the points in $S_\epsilon^m$, \textit{i.e.}
  $\Pi_\epsilon(A)=\frac{1}{|S_\epsilon^m|}\underset{\bm  \phi \in
    S_\epsilon^m}{\sum} \delta_{\bm  \phi}(A)$, with $|S_\epsilon^m|$ the
  cardinality of $S_\epsilon^m$.
\end{defn}

Loosely speaking the uniform distribution is hence defined as the
limit of a discrete uniform distribution on an equally spaced grid,
where the notion of ``equally spaced'' is determined by the distance
metric underlying $(M,d)$. Even though this definition is intuitive it
is not constructive. Apart from special cases, generating an
$\epsilon$-lattice is computationally difficult in a general metric space;
calculating the limit of $\epsilon$-lattices even more so. In addition it
is unclear, whether there is just one limit distribution all
$\epsilon$-lattices would converge to. To overcome these problems
\citet{demb:1990} uses the closely related notion of packing
numbers. The packing number $D(\epsilon,A,d)$ of a subset $A\subset M$ in
the metric $d$ is defined as the cardinality of an $\epsilon$-lattice on
$A$, and packing numbers are known for a number of metric spaces. An
$\epsilon$-pseudo-probability can then be defined as
$P_\epsilon(A)=\frac{D(\epsilon,A,d)}{D(\epsilon,M,d)}$.  It is straightforward to
see that $0\leq P_\epsilon(A) \leq 1$ and that $P_\epsilon(M)=1$, but packing
numbers are sub-additive and hence $P_\epsilon$ is not a probability
measure. However for disjoint sets $A'$ and $A''$ with minimum
distance $> \epsilon$ additivity holds, \textit{i.e.}  $P_\epsilon(A'\cup
A'')=P_\epsilon(A')+P_\epsilon(A'')$. \citet{demb:1990} then shows that
whenever $\underset{\epsilon \rightarrow 0}{\lim}\,P_\epsilon(A)$ exists for
any $A$, then the limit distribution is the unique uniform
distribution on $(M,d)$ (see \citet{demb:1990} or
\citet{ghos:ghos:rama:1997} for details). As packing numbers are
known for a number of metric spaces, this result provides a
constructive way for building uniform distributions, without the need
for explicitly constructing $\epsilon$-lattices.

Subsequently we consider the practically important case of a finite
number of parameters $p$ and assume that the metric of $(M,d)$, $d(\bm
\phi,\bm \phi_0)=d(\varphi(\bm \theta),\varphi(\bm \theta_0))=d^*(\bm
\theta,\bm \theta_0)$ in terms of $\bm \theta$ can be approximated by
a local quadratic approximation of the form
\begin{equation}
\label{eqn:quad}
d^*(\bm  \theta,\bm  \theta_0) = c_1\sqrt{c_2(\bm  \theta-\bm 
\theta_0)^T\bm  V(\bm  \theta_0)(\bm  \theta-\bm  \theta_0)+O(||\bm 
\theta-\bm  \theta_0||^k)},
\end{equation}
where $c_1,c_2>0$ are constants and $k\geq 3$. Equation
\eqref{eqn:quad} implies that $d(.,.)$ can locally be approximated by
a Euclidean metric. This is not a very strong condition, for a
sufficiently often differentiable metric $d(.,\bm \theta_0)$ one can
make use of a Taylor expansion of second order of $d(.,\bm
\theta_0)^2$ and apply the square root to obtain \eqref{eqn:quad}. The
following theorem calculates the distribution induced on $\bm \theta$
by imposing a uniform distribution in $(M,d)$, when assumption
\eqref{eqn:quad} holds. The proof is only a slight adaption of earlier
results by \citet{ghos:ghos:rama:1997}, see Appendix A.

\begin{thm}
  \label{thm:res}
  For a metric space $(M,d)$ and a bijective function $\varphi$,
  fulfilling \eqref{eqn:quad}, where $\bm  V(\bm  \theta)$ is a
  symmetric matrix with finite strictly positive eigenvalues $\forall
  \bm  \theta \in \bm  \Theta$ and continuous as a function of $\bm 
  \theta$, $P_\epsilon(A)=\frac{D(\epsilon,A,d)}{D(\epsilon,M,d)}$ for $A\subset
  \bm  \Theta$ converges to
  $$\frac{\int_{A}\sqrt{\det(\bm  V(\bm  \theta))}d\bm  \theta}{\int_{\bm 
      \Theta}\sqrt{\det(\bm  V(\bm  \theta))}d\bm  \theta},\; \mathit{as}\; \epsilon
  \rightarrow 0.$$ The density of the
  uniform probability distribution is hence given by:
  $$p(\bm  \theta) = \frac{\sqrt{\det(\bm  V(\bm  \theta))}}{\int_{\bm 
      \Theta}\sqrt{\det(\bm  V(\bm  \theta))}d\bm  \theta}.$$
\end{thm}

We note that the last result can be obtained as well by using
considerations based on Riemannian manifolds, in which case
(\ref{eqn:quad}) would be the Riemannian metric: For example
\citet{penn:2006} explicitly considers uniform distributions on
Riemannian manifolds and obtains the same result. We concentrated on
Dembski's derivation as it seems both more general and intuitive.

It is important to note that the so defined distribution is
independent of the parametrization. This is intuitively clear, as the
space $(M,d)$, where the uniform distribution is imposed, is fixed, no
matter, which parametrization is used. We illustrate this invariance
property for the special case of a Taylor approximation in the Theorem
below; for a proof see Appendix B.

\begin{thm}
  \label{thm:invar}
  Assume $(M,d)$ with $d(\bm \theta,\bm \theta_0)^2 = \frac{1}{2}(\bm
  \theta-\bm \theta_0)'\bm V(\bm \theta_0)(\bm \theta-\bm \theta_0) +
  O(||\bm \theta-\bm \theta_0||^3)$, where $\bm V(\bm
  \theta)=\left(\frac{d^2(\bm \theta,\bm \theta_0)}{\partial
      \theta_i\partial \theta_j}\right)_{i,j}$ evaluated at $\bm
  \theta$, which leads to a prior $p(\bm
  \theta)\propto \sqrt{\det(\bm V(\bm \theta))}$.\\
  When calculating the uniform distribution associated to the
  transformed parameter $g(\bm \theta)=\bm \gamma$, with $g:
  \mathbb{R}^p\rightarrow \mathbb{R}^p$ a bijective twice
  differentiable transformation, one obtains $p(\bm \gamma) \propto
  \det(\bm H(\bm \gamma))\sqrt{\det(\bm V(h(\bm \gamma)))}$, where $h$
  is the inverse of $g$ and $\bm H(\bm
  \gamma)=(\frac{\partial}{\partial \theta_1}h(\bm \theta), \ldots,
  \frac{\partial}{\partial \theta_p}h(\bm \theta))$ is the Jacobian
  matrix associated with $h$, which is the same result as applying the
  change of variables theorem to $p(\bm \theta)$.
\end{thm}
A technical restriction of the theory described in this section is the
concentration on compact metric spaces $\bm \Theta$. However, it is
possible to extend this based on taking limits of a sequence of
growing compact spaces, see the works of \citet{demb:1990} and
\citet{ghos:ghos:rama:1997} for details. Note that the resulting
limiting density does not need to be integrable.

\subsubsection{Examples}
\label{sec:examples}

\textbf{Non-functional uniform priors}

While the approach outlined in Section \ref{sec:general-approach} is
developed for general metric spaces, it coincides with standard
results about change of variables, when the metric space $M$ is a
compact subset of $\mathbb{R}^p$ as well. Suppose one would like to
use a uniform distribution for $\varphi(\bm \theta)$ with $\varphi(\bm
\theta): \mathbb{R}^p \rightarrow \mathbb{R}^p$ a bijective,
continuously differentiable function and then back-transform to $\bm
\theta$ scale. Using the standard change of variables theorem one
obtains: $p(\bm \theta)\propto|\det(D(\bm \theta))|$, where $D(\bm
\theta) = (\frac{\partial}{\partial \theta_1}\varphi(\bm \theta),
\ldots, \frac{\partial}{\partial \theta_p}\varphi(\bm \theta))$ is the
Jacobian matrix of the transformation $\varphi(\bm \theta)$. Framed in
the approach of the last section, the metric space $M$ is a compact
subset of $\mathbb{R}^p$ with the Euclidean metric in the transformed
space $d(\varphi(\bm \theta), \varphi(\bm \theta_0)) =
\sqrt{(\varphi(\bm \theta)-\varphi(\bm \theta_0))^T(\varphi(\bm
  \theta)-\varphi(\bm \theta_0))}$. A local linear approximation to
$\varphi(\bm \theta)-\varphi(\bm \theta_0)$ is $D(\bm \theta_0)(\bm
\theta-\bm \theta_0)$ with remainder $O(||\bm \theta-\bm
\theta_0||^2)$. Hence, one obtains $d(\varphi(\bm \theta), \varphi(\bm
\theta_0)) = \sqrt{(\bm \theta-\bm \theta_0)^TD(\bm \theta)^TD(\bm
  \theta)(\bm \theta-\bm \theta_0)+O(||\bm \theta-\bm
  \theta_0||^3)}$. Applying Theorem \ref{thm:res} one ends up with the
desired distribution $p(\bm \theta) \propto \sqrt{\det(D(\bm
  \theta)^TD(\bm \theta))}=|\det(D(\bm \theta))|$.

\textbf{Jeffreys Prior}

Another special case of this general approach is Jeffreys prior
itself. \citet{jeff:1961} described his rule by noting that
\eqref{eqn:quad} approximates the empirical Hellinger distance (as
well as the empirical Kullback-Leibler divergence) between the
residual distributions in a statistical model, when $\bm V(\bm
\theta)$ is the Fisher information matrix.  In this situation the
parameters of a statistical model $\bm \theta$ are mapped into the
space of residual densities and this space is used to define the
notion of distance between the $\bm \theta$'s. Applying the machinery
from the last section then leads to a uniform distribution on the
space of residual densities. This interpretation of Jeffreys rule is
rare, but has been noted among others for example by
\citet[ch. 3.6]{kass:wass:1996}. \citet{ghos:ghos:rama:1997} and
\citet{bala:1997} explicitly derive Jeffreys rule from these
principles. From this viewpoint Jeffreys prior is hence useful as a
universal ``default'' prior, because it gives equal weights to all
possible residual densities underlying a statistical model. However,
the used metric can depend, for example, on values of covariates,
which is undesirable in the nonlinear regression application, as
discussed in the introduction.

\textbf{Triangular Distribution}

In this example Definition \ref{eqn:def} is directly used to
numerically approximate a uniform distribution on a metric space. This
is can be done in the case $p=1$, where the construction of
$\epsilon-$lattices is easily possible numerically.

\begin{figure}
  \centerline{\includegraphics[width=0.9\textwidth]{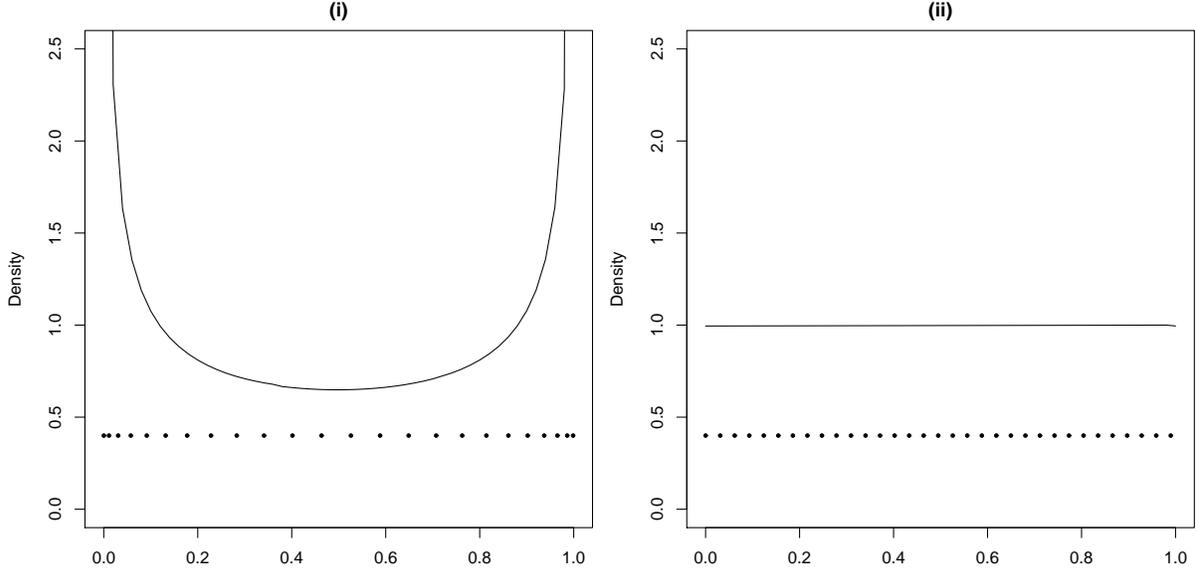}}
  \caption{Numerically calculated uniform distributions in the (i)
    Hellinger metric and (ii) the Kolmogorov metric. The solid curves
    are based on interpolation of the empirical distribution functions
    of 0.005-lattices followed by differentiation, the dots represent
    an 0.03-lattice.}
 \label{fig:triang}
\end{figure}

The triangular distribution, with density
\begin{equation}
  \label{eq:triang}
  p(x|\theta)=\begin{cases} 2x/\theta, & 0 < x \leq
    \theta\\ 2(1-x)/(1-\theta), & \theta \leq x < 1 \end{cases},
\end{equation}
for $\theta \in (0,1)$ is a simple, yet versatile distribution, for
which the Jeffreys prior does not exist
\citep{berg:bern:sun:2009}. One possible metric space where to impose
the uniform distribution is the space of triangular densities or
triangular distribution functions parametrized by $\theta$. Several
metrics might be used, we will consider the Hellinger metric
$d_H(\theta_1, \theta_2)=\left\{\int_0^1
  \left(\sqrt{p(x|\theta_1)}-\sqrt{p(x|\theta_2)}\right)^2dx\right\}^\frac{1}{2}$
and the Kolmogorov metric $d_K(\theta_1, \theta_2) = \sup_{y\in [0,1]}
|\int_0^y(p(x|\theta_1)-p(x|\theta_2))dx|$. Numerically calculating
the corresponding $\epsilon-$lattices one obtains the distributions
displayed in Figure \ref{fig:triang}. Interestingly one can observe
that the calculated uniform distribution in the Hellinger metric space
is equal to a $Beta(1/2, 1/2)$ distribution (as the reference prior in
\citet{berg:bern:sun:2009}), while the calculated functional uniform
distribution in the Kolmogorov metric results in a uniform
distribution on $[0,1]$.

\subsection{Nonlinear Regression}
\label{sec:nonreg}

The implicit assumption when employing a nonlinear regression function
$\mu(x, \bm \theta)$, is that for one $\bm \theta$ the shape of the
function $\mu(x, \bm \theta)$ will adequately describe reality. It is
usually unclear, however, which of these shapes is the right one. A
uniform distribution on the functional \textit{shapes} hence seems to
be a reasonable prior. A suited metric space is consequently the space
of functions $\mu(., \bm \theta)$, with $x\in
\mathcal{X}\subset\mathbb{R}$, $\bm \theta \in K \subset \mathbb{R}^p$
with compact $K$ and metric for example given by the $L_2$ distance
$d(\bm \theta,\bm \theta_0)=\sqrt{\int_{\mathcal X}(\mu(x,\bm
  \theta)-\mu(x,\bm \theta_0))^2dx}$. By a first order Taylor
expansion one obtains $\mu(x, \bm \theta)-\mu(x, \bm \theta_0)=J_x(\bm
\theta_0)(\bm \theta - \bm \theta_0) + O(||\bm \theta - \bm
\theta_0||^2)$, where $J_x(\bm \theta_0)=\frac{\partial}{\partial \bm
  \theta} \mu(x, \bm \theta)$ is the row vector of first partial
derivatives. This results in an approximation of form $(\mu(x,\bm
\theta)-\mu(x,\bm \theta_0))^2=(\bm \theta - \bm \theta_0)^TJ_x(\bm
\theta_0)^TJ_x(\bm \theta_0)(\bm \theta - \bm \theta_0)+O(||\bm \theta
- \bm \theta_0||^3)$. Integrating this with respect to $x$ and taking
the square root, leads to an approximation of $d(\bm \theta,\bm
\theta_0)$ of form $\sqrt{(\bm \theta-\bm \theta_0)^T\bm Z^*(\bm
  \theta_0)(\bm \theta-\bm \theta_0)+O(||\bm \theta-\bm
  \theta_0||^3)},$ where $\bm Z^*(\bm
\theta)=\int_{\mathcal{X}}J_x(\bm \theta)^TJ_x(\bm
\theta)dx$. Consequently, from Theorem \ref{thm:res} the functional
uniform distribution for $\bm \theta$ equals $p(\bm \theta) \propto
\sqrt{\det(\bm Z^*(\bm \theta))}.$ In the special case of a linear
model $\mu(x,\bm \theta)=f(x)^T\bm \theta$, the functional uniform
distribution collapses to a constant prior distribution, which is the
uniform distribution on $\bm \Theta$ for compact $\bm \Theta$ and
improper, when extending to non-compact $\bm \Theta$.

\begin{figure}
  \centerline{\includegraphics[width=0.9\textwidth]{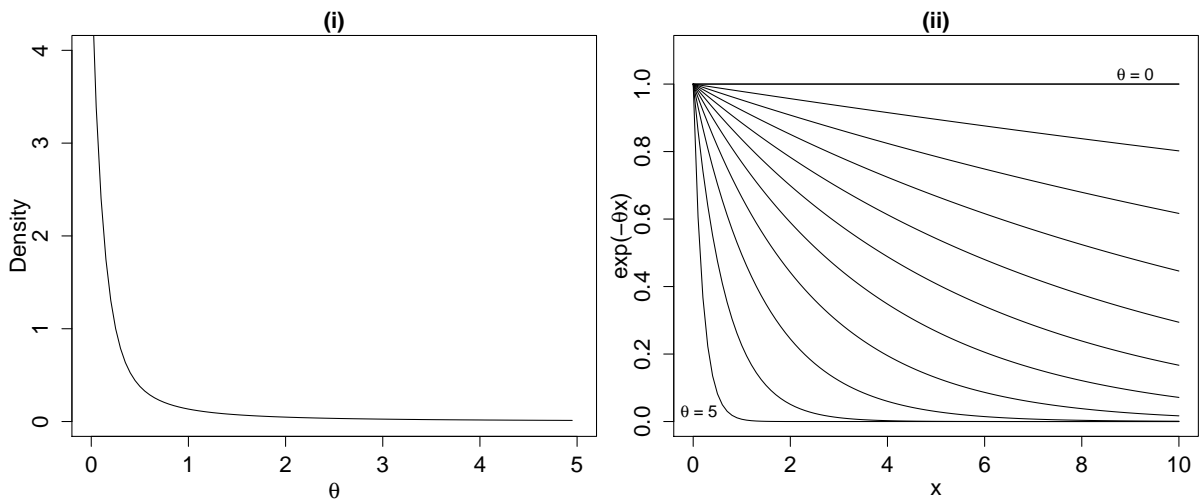}}
 \caption{(i) Display of the functional uniform distribution on $\theta$ scale;
   (ii) Display of the regression function $\exp(-\theta x)$
   for $\theta=0$, $\theta = 5$ and the $\theta$ corresponding to the
   $i/10$ quantile $i=1,\ldots,9$ of the functional uniform
   distribution.}
 \label{fig:expo2}
\end{figure}

We now revisit the exponential regression example from the
introduction.  In this case one obtains $J_x(\theta)=-x\exp(-\theta
x)$, calculating $\int_0^{10}J_x(\theta)^2dx$ and applying the square
root, one obtains $p(\theta)\propto
\exp(-10\theta)\sqrt{\frac{\exp(20\theta)-200\theta^2-20\theta-1}{\theta^3}}$,
normalizing this leads to the prior displayed in Figure
\ref{fig:expo2} (i). On the $\theta$ scale the shape based functional
uniform density hence leads to a rather non-uniform distribution. In
Figure \ref{fig:expo2} (ii) one can observe that the probability mass
is distributed uniformly over the different shapes, as desired.

An advantage of the functional uniform prior over the uniform prior is
that it is independent of the choice of parameterization and not
particularly sensitive to the potential choice of the bounds, provided
all major shapes of the underlying function are covered. In Figure
\ref{fig:expo2} (i) one can see that the density is already rather
small at $\theta=5$ as most of the underlying functional shapes are
already covered. In fact in this example one can extend the functional
uniform distribution from the compact interval $[0,5]$ to a proper
distribution on $[0, \infty)$.

Although the choice of the $L_2$ metric for $d$ seems reasonable in a
variety of situations, other choices are possible. One could for
example use a weighted version of the $L_2$ distance, when interest is
in particular regions of the design space $\mathcal{X}$. In fact
Jeffreys prior can be identified as a special case, when the assumed
residual model is given by a homoscedastic normal distribution. In
this situation the empirical measure on the design points is used as a
weighting measure.  The Jeffreys prior has also been mentioned by
\citet[p. 217]{bate:watts:1988}, as a prior that is uniform on the
response surfaces, but the possibility of an alternative weighting
measures has not been considered.

One potential obstacle in the use of the proposed functional uniform
prior is the fact that it can be computationally challenging to
calculate. In some of the situations it might be possible to calculate
$\bm Z^*(\bm \theta)$ analytically in others one might need to use
numerical integration to approximate the underlying
integrals. However, it needs to be noted that the prior only needs to
be calculated once, as the prior is independent of the observed data
(it only depends on the design region $\mathcal{X}$ and on potential
parameter bounds), and can then be approximated for example in terms
of more commonly used distributions. This approximation can then be
reused in different modelling situations.

\section{Applications}
\label{sec:appl}

In this section, we will evaluate the proposed functional uniform
priors for nonlinear regression. One application of nonlinear
regression is in the context of pharmaceutical dose-finding trials. A
challenge in these trials is that the variability in the response is
usually large and the number of used doses fairly small, so that the
underlying inference problem is challenging, despite an often
seemingly large sample size. The priors will first be tested in a real
example, then the frequentist operating characteristics of the
proposed functional uniform priors are assessed more formally in a
simulation study for a binary endpoint. In the last example we will
use the functional uniform distribution for calculation of a Bayesian
optimal design in the exponential regression example.

\subsection{Irritable Bowel Syndrome Dose-Response Study}
\label{sec:df-regression}

Here the \texttt{IBScovars} data set taken from the
\texttt{DoseFinding} package will be used \citet{DoseFinding}. The
data were part of a dose ranging trial on a compound for the treatment
of the irritable bowel syndrome with four active doses 1, 2, 3, 4
equally distributed in the dose range $[0, 4]$ and placebo.  The
primary endpoint was a baseline adjusted abdominal pain score with
larger values corresponding to a better treatment effect. In total 369
patients completed the study, with nearly balanced allocation across
the doses. Assume a normal distribution is used to model the residual
error and that the hyperbolic Emax model $\mu(x,\bm \theta) =
\theta_0+\theta_1x/(\theta_2+x)$ was chosen to describe the
dose-response relationship. The parameters $\theta_0$ and $\theta_1$
determine the placebo mean and the asymptotic maximum effect, while
the parameter $\theta_2$ determines the dose that gives 50 percent of
the asymptotic maximum effect, so that it determines the steepness of
the curve.  In clinical practice vague prior information typically
exists for $\theta_0$ and $\theta_1$, but for illustration here we use
improper constant priors for these two parameters and a prior
proportional to $\sigma^{-2}$ for $\sigma$. For the nonlinear
parameter $\theta_2$ we will use a uniform prior and the functional
uniform prior distribution. When using a uniform distribution for
$\theta_2$ it is necessary to assume bounds, as otherwise an improper
posterior distribution may arise. We will use the bounds $[0.004,6]$
here, the selection of the boundaries is based on the fact that
practically all of the shapes of the underlying model are covered
taking into account that the dose range is $[0,4]$. For comparability
for the functional uniform prior the same bounds were used, although
one can extend it to an integrable density on $[0,\infty)$.  The
functional uniform prior will be used based on the function space
defined by $x/(\theta_2+x)$. Performing the calculations described in
Section \ref{sec:nonreg} one obtains
$J^2_x={{x^2}/{\left(x+\theta_2\right)^4}}$, calculating the integral
$\int_0^4J^2_x(\theta_2)\,dx$ and applying the square root leads to
$p(\theta_2)\propto
{1}/{\sqrt{\theta_2^4+12\theta_2^3+48\theta_2^2+64\theta_2}}1_{[0.004,6]}(\theta_2)$. Similar
to the exponential regression example in the introduction, a uniform
distribution on $\theta_2$ space induces an informative distribution
in the space of functional shapes. Shapes corresponding to larger
values of $\theta_2$ (say $>3$) correspond to almost linear shapes,
while only very small values of $\theta_2$ lead to more pronounced
concave shapes. A uniform prior on $\theta_2$ hence induces a prior
that favors linear shapes over steeply increasing model shapes.

\begin{figure}
  \centerline{\includegraphics[width=0.9\textwidth]{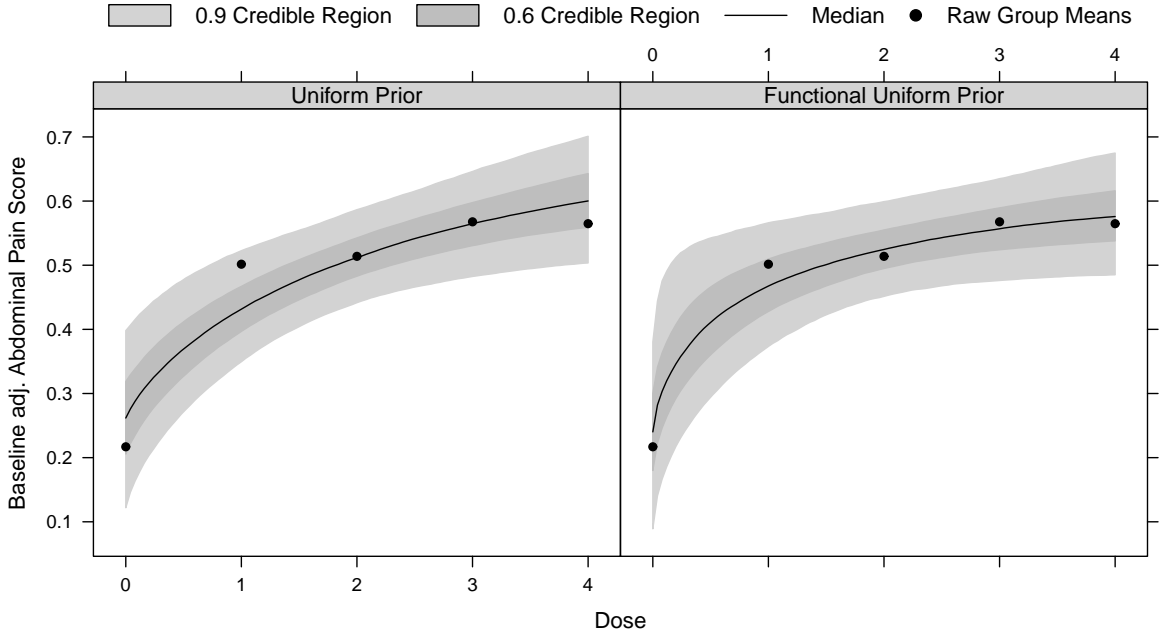}}
 \caption{Posterior the dose-response curve under a uniform and a
   functional uniform prior for $\theta_2$.}
 \label{fig:IBS}
\end{figure}

We used importance sampling resampling based on a proposal
distribution generated by the iterated Laplace approximation to
implement the model (see, \citet{born:2011}). In Figure \ref{fig:IBS}
one can observe the posterior uncertainty intervals under the two
prior distributions. As is visible, the bias towards linear shapes,
when using a uniform distribution for $\theta_2$ pertains in the
posterior distribution. This happens despite the rather large sample
size, and despite the fact that the response at the doses 0, 4 and
particularly at dose 1 are not very well fitted by a linear shape. So
the posterior seems to be rather sensitive to the prior uniform
distribution. The posterior based on the shape based functional
uniform prior, in contrast, fits the data better at all doses, and
seems to provide a more realistic measure of uncertainty for the
dose-response curve, particularly for $x\in (0,1)$.

\subsection{Simulations}
\label{sec:simulations}

One might expect that the functional uniform prior distribution works
acceptable no matter which functional shape is the true one. To
investigate this in more detail and to compare this prior to other
prior distributions in terms of their frequentist performance,
simulation studies have been conducted. Here we report results from
simulations in the context of binary nonlinear regression.

For simulation the power model $\mu(x,
\theta)=\theta_0+\theta_1x^{\theta_2}$ will be used to model the
response probability depending on $x$. The parameters $\theta_0$ and
$\theta_1$ are hence subject to $\theta_0+\theta_1\leq 1$ and
$\theta_0,\theta_1\geq 0$, as a probability is modelled. Note that
only $\theta_2$ enters the model function non-linearly

The doses $0, 0.05, 0.2, 0.6, 1$ are to be used with equal allocations
of 20 patients per dose. We use four scenarios in this case: in the
first three cases the power model is used with $\theta_0=0.2$ and
$\theta_1=0.6$, while $\theta_2$ is equal to $0.4$ (Power 1), $1$
(Linear) and $4$ (Power 2). In addition we provide one scenario, where
an Emax model $0.2+0.6/(x+0.05)$ is the truth. The Emax scenario is
added to investigate the behaviour under misspecification of the
model. Each simulation scenario will be repeated 1000 times.

We will compare the functional uniform prior distribution to the
uniform distribution on the parameters and to the Jeffreys prior
distribution. For the uniform prior distribution approach uniform
prior distributions were assumed for all parameters, and the nonlinear
parameter $\theta_2$ was assumed to be within $[0.05, 20]$ to ensure
integrability. The same bounds are used for the two other approaches
for comparability. For the functional uniform prior approach, uniform
priors are used for $\theta_0$ and $\theta_1$, while for
$\theta_2$ the functional uniform prior will be used on the function
space defined by $x^{\theta_2}$. The prior can be calculated to be
$p(\theta_2)\propto
1/\sqrt{2\theta_2^3/3+\theta_2^2+\theta_2/2+1}$. For the Jeffreys
prior approach we used a prior proportional to $\sqrt{{\det(I(\bm
    \theta))}}$, within the imposed parameter bounds.  For analysis we
used MCMC based on the HITRO algorithm, which is an MCMC sampler that
combines the hit and run algorithm with the ratio of uniforms
transformation. It does not need tuning and is hence well suited for a
simulation study. The sampler is implemented in the \texttt{Runuran}
package \citep{leyd:hoer:2010}, computations were performed with
\texttt{R} \citep{R}. 10000 MCMC samples are used from the
corresponding posterior distributions in each case, using a burnin
phase of 1000 a thinning of 2.

\begin{table}
\begin{center}
\begin{tabular}{|c|c|cccc|}
  \hline
  Prior & Model & MAE$_1$ & MAE$_2$ &CP & ILE \\
  \hline
  Uniform     & Linear & 0.082 & 0.062 & 0.819 & 0.259  \\ 
              & Power 1& 0.079 & 0.056 & 0.816 & 0.255  \\ 
              & Power 2& 0.066 & 0.067 & 0.881 & 0.220  \\ 
              & Emax   & 0.073 & 0.056 & 0.780 & 0.226  \\ \hline
  Jeffreys    & Linear & 0.058 & 0.065 & 0.900 & 0.233  \\ 
              & Power 1& 0.054 & 0.056 & 0.901 & 0.220  \\ 
              & Power 2& 0.056 & 0.073 & 0.895 & 0.227  \\ 
              & Emax   & 0.056 & 0.055 & 0.845 & 0.196  \\ \hline
  Func. Unif. & Linear & 0.060 & 0.060 & 0.892 & 0.240  \\ 
              & Power 1& 0.057 & 0.053 & 0.893 & 0.226 \\ 
              & Power 2& 0.057 & 0.070 & 0.912 & 0.240 \\ 
              & Emax   & 0.059 & 0.054 & 0.834 & 0.203 \\ 
   \hline
\end{tabular}

\caption{Estimation of dose-response; MAE$_1$ and MAE$_2$ correspond to the
  dose-response estimation error (for the posterior median and mode), CP
  denotes the average coverage probability of pointwise 0.9
  credibility intervals and ILE denotes the average credibility
  interval lengths.}
  \label{tab:results4}
\end{center}
\end{table}

In Table \ref{tab:results4} one can observe the estimation results in
terms of the mean absolute estimation error for the dose-response
function, $\mathrm{MAE} = 1/9\sum_{i=0}^8|\mu(i/8)-\hat{\mu}(i/8)|$,
where $\mu(.)$ is the underlying true function and $\hat{\mu}(.)$ is
either the point wise posterior median (corresponding to MAE$_1$) or
the prediction corresponding to the posterior mode for the parameters
(MAE$_2$), the posterior mode for the uniform prior is equal to the
maximum likelihood estimate. The values displayed in Table
\ref{tab:results4} are the average $\mathrm{MAE}$ over 1000
repetitions. In addition for each simulation the 0.9 credibility
intervals at the dose-levels $0,1/8,2/8,...,1$ have been
calculated. The number given in the Table is $CP = 1/9\sum_{i=0}^8
\hat{P}_{i/8}$, where $\hat{P}_{d}$ is the average coverage
probability of the 0.9 credibility interval at dose $d$ over 1000
simulation runs. In addition the average length of the credibility
intervals has been calculated as $ILE= 1/9\sum_{i=0}^8 \hat{L}_{i/8}$,
where $L_d$ is the average length of the 0.9 credibility interval at
dose $d$ over 1000 simulation runs.  For estimation of the
dose-response Jeffreys prior and the functional uniform prior improve
upon the uniform prior distribution, while the Jeffreys prior and the
functional uniform prior are close, with slight advantages for the
Jeffreys prior. In terms of the credibility intervals the functional
uniform and Jeffreys prior roughly keep their nominal level for the
linear, and the power model cases, while the uniform prior probability
does not. None of the priors achieves the nominal level for the Emax
model, which is probably due to the fact that the Emax model is too
different from the power model. Interestingly the credibility
intervals of the uniform prior are larger than those of the other two
priors, but lead to a smaller coverage probability.

Table \ref{tab:results2} provides the estimation results with respect
to parameter estimation. The main message here is that all priors
perform roughly equal for estimation of the linear parameters
$\theta_0$ and $\theta_1$. For the nonlinear parameters, Jeffreys
prior distribution and the functional uniform prior perform better
than the uniform disribution.

In summary the functional uniform prior hence performs roughly equally
well as the Jeffreys prior in these simulations. However, the
functional uniform prior has the pragmatic and conceptual advantages
that it does not depend on the observed covariates, and can thus be
used for example for calculation of a Bayesian optimal design, or in
sequential situations.

\begin{table}
  \centering
  {\small
  \begin{tabular}{|c|c|cccc|cccc|} \hline
    \multicolumn{2}{|c|}{}  & \multicolumn{4}{c|}{Uniform Prior} &
    \multicolumn{4}{c|}{Functional Uniform Prior} \\ \hline
    Scenario    & $N$ & MAE$_1$ & MAE$_2$ & CP & ILE & MAE$_1$ & MAE$_2$ & CP & ILE\\ \hline
    Sig. Emax 1 & 125 & 0.256 & 0.277 & 0.903 & 1.098 & 0.230 & 0.270 & 0.914 & 1.028\\
    Sig. Emax 2 &     & 0.278 & 0.283 & 0.895 & 1.144 & 0.258 & 0.278 & 0.909 & 1.089\\
    Sig. Emax 3 &     & 0.243 & 0.275 & 0.902 & 1.014 & 0.251 & 0.262 & 0.898 & 1.030\\
    Linear      &     & 0.266 & 0.291 & 0.901 & 1.100 & 0.241 & 0.289 & 0.918 & 1.057\\
    Quadratic   &     & 0.272 & 0.278 & 0.880 & 1.109 & 0.242 & 0.276 & 0.898 & 1.038\\ \hline
    Sig. Emax 1 & 250 & 0.185 & 0.214 & 0.908 & 0.818 & 0.167 & 0.209 & 0.920 & 0.768\\
    Sig. Emax 2 &     & 0.196 & 0.206 & 0.908 & 0.850 & 0.187 & 0.201 & 0.910 & 0.811\\
    Sig. Emax 3 &     & 0.174 & 0.202 & 0.913 & 0.738 & 0.170 & 0.188 & 0.912 & 0.744\\
    Linear      &     & 0.200 & 0.209 & 0.891 & 0.831 & 0.189 & 0.211 & 0.900 & 0.794\\
    Quadratic   &     & 0.201 & 0.215 & 0.881 & 0.839 & 0.185 & 0.216 & 0.886 & 0.782\\\hline
  \end{tabular}
  \caption{Estimation of dose-response; MAE$_1$ and MAE$_2$ correspond
    to the estimation error at the doses 0,1,...,8 (for the posterior
    median and the posterior mode), CP denotes the average
    coverage probability of pointwise 0.9 credibility intervals and ILE
    denotes the average credibility interval lengths.}
  \label{tab:results2}
}
\end{table}

\subsection{Bayesian optimal design for exponential regression}
\label{sec:optdes}

In this section we will use the prior distribution for the exponential
regression model derived in Section \ref{sec:nonreg} to calculate a
Bayesian optimal design. When assuming a homoscedastic normal model,
the Fisher information is $I(d,\theta)\propto \sum w_i x_i^2
\exp(-2\theta x_i)$.  Hence minimizing $-\log(I(d,\theta))$ will lead
to a design with most information. Unfortunately the expression
depends on $\theta$, which is of course unknown before the
experiment. One way of dealing with this uncertainty are Bayesian
optimal designs, where one optimizes the design criterion averaged
with respect to a prior distribution: $-\int \log(I(d,\theta))
p(\theta)d\theta$. In this situation we will use the uniform and
functional uniform prior distribution (see Figures \ref{fig:expo1} and
\ref{fig:expo2}) both on the interval $[0,5]$ for calculation of the
optimal design. Restricting the design space to $x \in [0,10]$ and
only performing the optimization up to 5 design points, one ends up
with the weights $\bm w=(0.956,0.022,0.022)$ on the design points
$x=(0.38,4.04,10)$, for the uniform prior, while the functional
uniform prior distribution leads to a design of the form $\bm
w=(0.19,0.3,0.51)$ and $x=(0.54,2.35,10)$. The design corresponding to
the functional uniform prior hence spreads its allocation weights more
uniformly on the design range, whereas the uniform prior results in
essentially one major design point.

\begin{figure}
  \centerline{\includegraphics[width=0.9\textwidth]{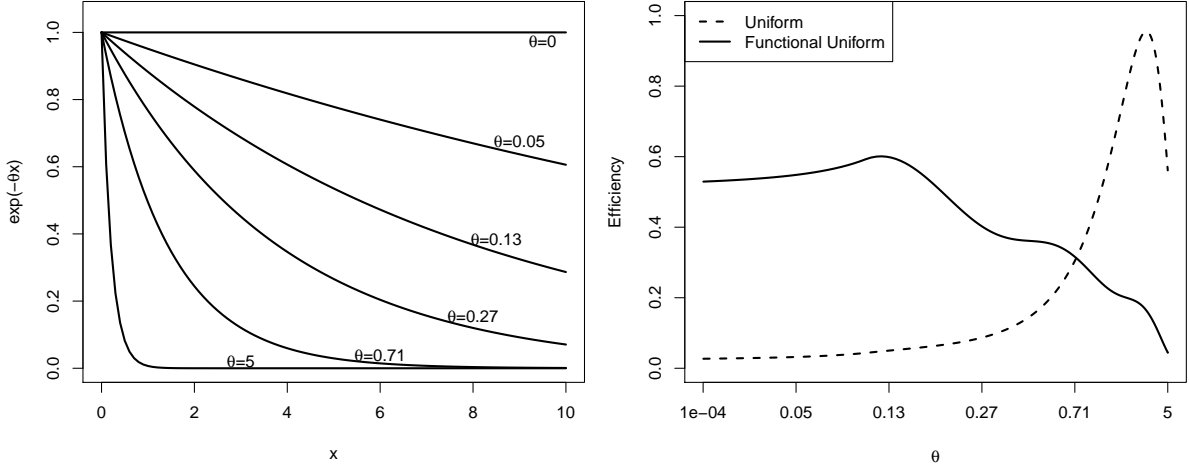}}
  \caption{Efficiency of the two designs for different shapes.}
 \label{fig:shapeEff}
\end{figure}

One way of comparing the two calculated designs is to look at the
efficiency $\mathrm{Eff}(d, \theta)=\exp(\log(I(d,
\theta))-\log(I(d_{opt}(\theta), \theta)))$, of the calculated
designs, with respect to the design $d_{opt}(\theta)$ that is locally
optimal for the parameter value $\theta$, for a range of different
shapes. In Figure \ref{fig:shapeEff} we plot the efficiency for the
different shapes on the functional shape scale.

One can observe that the uniform prior design is only efficient for
the sharply decreasing shapes with $\theta > 0.71$, but otherwise has
very low efficiency.  The functional uniform prior improves quite a
bit over the uniform prior distribution for most of the functional
shape space, and provides at least a reasonable efficiency for most
shapes.

\section{Conclusions}
\label{sec:conclusions}

A main motivation for this work is the practical limitation of the
classical Jeffreys prior that it cannot be used in nonlinear
regression settings, where the prior needs to be specified before data
collection, for example when one wants to calculate a Bayesian optimal
design or in adaptive dose-finding trials. For this purpose the
functional uniform distribution has been introduced, which imposes a
distribution on the parameters, so that it is uniform in the
functional shapes underlying the nonlinear regression function. This
was achieved by using a general framework for constructing uniform
distributions based on earlier work by \citet{demb:1990} and
\citet{ghos:ghos:rama:1997}. We investigated the functional uniform
prior for nonlinear regression in a real example, a simulation study
and an optimal design problem where it showed very satisfactory
performance. 

There is no reason to call the priors proposed in this article
globally uninformative, because one needs to choose the space $M$ and
in particular the metric $d$, where to impose the uniform
distribution. The priors derived from the theory in Section
\ref{sec:general-approach} might then be considered uninformative in
the particular aspect that $(M,d)$ reflects.  In the case of nonlinear
regression we argue that the uniform distribution on the space of
functional shapes is often, depending of course on the considered
application, a \textit{reasonable} assumption for nonlinear regression
when particular prior information is lacking. However, this also does
not apply generally: A situation, where the functional uniform prior
might not be adequate, occurs, for example, when the considered
nonlinear model is extremely flexible, containing virtually all
continuous functions for example (as neural network models). In this
case it is often more adequate to concentrate most prior probability
on a reasonable subset of the function space (\textit{e.g.}, smooth
functions), rather than building a uniform distribution on all
potential shapes, including shapes that might be implausible a-priori.

The theory outlined in Section \ref{sec:methodology} might be of
interest to formulate functional uniform priors also for other type of
models with a nonlinear aspect. In quite a few modelling situations
one might be able to find a space $(M,d)$, where imposing a uniform
distribution is plausible and then back-transform this distribution to
the parameter scale.
\citet{ghos:ghos:rama:1997} employ this idea, when $(M,d)$ is a
space of densities and define priors for nonparametric density
estimation. Another application could be the estimation of covariance
matrices: \citet{dryd:kolo:zhou:2009} discuss the use of more
adequate non-Euclidean distance metrics for covariance matrices, which
would in our framework define the metric space for imposing the
uniform distribution. \citet{paul:2005} derives default priors
for Gaussian process interpolation, which are rather time consuming to
evaluate. In this situation might choose the space of the covariance
functions as $(M,d)$.

\begin{appendix}

\section{Proof of Theorem 1}
\label{sec:the1}

\citet{ghos:ghos:rama:1997} prove a closely related result, when the
underlying metric is the Hellinger distance and $M$ the space of
residual densities. We review their proof and adapt to metrics of the
form (1) and proceed in two parts. Part A summarizes the
proof of \citet{ghos:ghos:rama:1997} for completeness and part B
provides additional Lemmas needed in our situation.

\textit{Part A}\\
The proof starts by covering $\bm \Theta$ with hypercubes and inner
hypercubes placed inside these cubes. Let $A_1,\ldots,A_J$ be the
intersections of $A$ with the hypercubes and $A'_1,\ldots,A'_J$ be the
intersections of $A$ with the inner hypercubes. Now separate the
hypercubes and inner hypercubes so that each inner hypercube is at
least $\epsilon$ apart from any other in the $d(.,.)$ metric (this is
possible, when the results proved in Lemma 1 hold). By the
sub-additivity of packing numbers one then has
$\sum_jD(\epsilon,A'_j,d)\leq D(\epsilon, A, d)\leq \sum_jD(\epsilon,
A_j,d)$ and $\sum_jD(\epsilon,\bm \Theta'_j,d)\leq D(\epsilon, \bm
\Theta, d)\leq \sum_jD(\epsilon, \bm \Theta_j,d)$, where $\bm
\Theta'_j$ and $\bm \Theta_j$ denote the intersection with the
hypercubes and inner hypercubes.

Now an upper an lower bound for $D(\epsilon, A_j,d)$ is derived based on
the local Euclidean approximation (1) to the metric
$d$. For a Euclidean metric one can calculate the packing number
explicitly, see \citet{kolm:tiho:1961}. Up to proportionality
$D(\epsilon,A,||.||)$ is given by $vol(A)\epsilon^{-p}$, consequently for a
metric of form $\sqrt{(\bm \theta-\bm \theta')^T\bm V(\bm \theta-\bm
  \theta')},$ with $\bm V$ a fixed positive definite matrix the
packing number is up to proportionality $\sqrt{\mathrm{det}(\bm
  V)}vol(A)\epsilon^{-p}$. Using the local Euclidean approximation
(1) and Lemma 2 one can derive lower and upper
bounds for $D(\epsilon, A_j,d)$ and $D(\epsilon, A'_j,d)$ in terms of
$\sqrt{\mathrm{det}(\bm V(\bm \theta_j))}vol(A_j)\epsilon^{-p}$ and
$\sqrt{\mathrm{det}(\bm V(\bm \theta_j))}vol(A'_j)\epsilon^{-p}$. and
similarly for $D(\epsilon,\bm \Theta,d)$ and thus for $P_\epsilon(A)= D(\epsilon,
A,d)/D(\epsilon, \bm \Theta,d)$. As the size of the hypercubes goes to
zero the bounds become sharper (see Lemma 2) and lower and
upper bound $P_\epsilon(A)$ converge to $\frac{\int_{A}\sqrt{\det(\bm
    V(\bm \theta))}d\bm \theta}{\int_{\bm \Theta}\sqrt{\det(\bm V(\bm
    \theta))}d\bm \theta}$, see \citet{ghos:ghos:rama:1997} for
the details of this argument.

\textit{Part B}\\
Without loss of generality we focus on setting $c_1=c_2=1$ in
(1) for what follows.

\begin{lem}
  \label{lemma2}
  For symmetric, positive definite $\bm V(\bm \theta)$ 
there exist $l^*,u^*>0$ for $\bm \theta, \bm \theta' \in \bm \Theta$ so that
$$l^*||\bm \theta-\bm \theta'|| \leq d(\bm \theta,\bm \theta') \leq u^*||\bm \theta-\bm \theta'||$$
\end{lem}

Proof:\\
$$\frac{d(\bm \theta, \bm \theta')^2}{||\bm \theta-\bm
  \theta'||^2}=\frac{{(\bm \theta-\bm \theta')^T\bm V(\bm \theta')(\bm
    \theta-\bm \theta')+O(||\bm \theta-\bm \theta'||^p)}}{||\bm
  \theta-\bm \theta'||^2}.$$ Now by an eigendecomposition and the
compactness of $\bm \Theta$ and continuity of $\bm V(\bm \theta)$ one
knows that there exist $l,u>0$ so that $$l(\bm \theta-\bm
\theta')^T(\bm \theta-\bm \theta') \leq (\bm \theta-\bm \theta')^T\bm
V(\bm \theta)(\bm \theta-\bm \theta')\leq u(\bm \theta-\bm
\theta')^T(\bm \theta-\bm \theta').$$ So that in total we get a lower
and upper bound by $u^{*2}=u+\kappa$, where $\kappa = \max(\max_{\bm
  \theta \in \bm \Theta} O(||\bm \theta-\bm \theta'||^{p-2}),0)$, and
similarly lower bounded. $\Box$

\begin{lem}
  \label{lemma3}
  For $\bm \theta, \bm \theta',\bm \theta^*$ lying in a
  hypercube $Q\subset \bm \Theta$ we obtain
$$k_1(\bm \theta-\bm \theta')^T\bm V(\bm \theta^*)(\bm \theta-\bm \theta')
\leq d^2(\bm \theta,\bm \theta') \leq k_2(\bm \theta-\bm \theta')^T\bm V(\bm
\theta^*)(\bm \theta-\bm \theta'),$$ where $k_1 \rightarrow c_0$ and
$k_2 \rightarrow c_0$ for $c_0 > 0$, when the side length of $Q$
converges to 0.
\end{lem}

Proof:
\begin{eqnarray*}
  \frac{d^2(\bm \theta, \bm \theta')}{(\bm \theta-\bm \theta')'\bm V(\bm \theta^*)(\bm \theta-\bm \theta')}
  &=& \frac{(\bm \theta-\bm \theta')^T\bm V(\bm \theta')(\bm \theta-\bm \theta')}{(\bm \theta-\bm \theta')^T\bm V(\bm \theta^*)(\bm \theta-\bm \theta')}
  +\frac{O(||\bm \theta-\bm \theta'||^p)}{(\bm \theta-\bm
    \theta')^T\bm V(\bm \theta^*)(\bm \theta-\bm \theta')} \nonumber\\
  &=& \frac{(\bm \theta-\bm \theta')^T\bm V(\bm \theta')(\bm \theta-\bm \theta')}{(\bm \theta-\bm \theta')^T\bm V(\bm \theta^*)(\bm \theta-\bm \theta')}
  +O(||\bm \theta-\bm \theta'||^{p-2}) \nonumber \\
  &=& 1+\frac{(\bm \theta-\bm \theta')^T(\bm V(\bm \theta')-\bm V(\bm \theta^*))(\bm \theta-\bm \theta')}{(\bm \theta-\bm \theta')^T\bm V(\bm \theta^*)(\bm \theta-\bm \theta')}+O(||\bm \theta-\bm \theta'||^{p-2})
\end{eqnarray*}

Now $\bm V(\bm \theta)$ is continuous, so one can lower and upper
bound the second summand on $Q$. It converges to zero and hence the
bounds towards each other when the size of the hypercubes shrinks (and
by this $\bm \theta^* \rightarrow \bm \theta'$). Upper and lower
bounding $O(||\bm \theta-\bm \theta'||^{p-2}$ implies the desired
result. $\Box$

\vspace{0.5cm}

\section{Proof of Theorem 2}
\label{sec:the2}

Consider the distance metric $d^*(\bm \gamma, \bm \gamma_0)=d(h(\bm
\gamma),h(\bm \gamma_0))$. To show invariance of the proposed
procedure, the uniform distribution derived from $d^*$ needs to be
$p(\bm \gamma) \propto \det(\bm H(\bm \gamma))\sqrt{\det(\bm V(h(\bm
  \gamma)))}$, which is the distribution derived from $p(\bm \theta)
\propto \sqrt{\det(\bm V(\bm \theta))}$ using a change of variables.

A second order Taylor expansion of $d^2(h(\bm \gamma),h(\bm
\gamma_0))$ in $\bm \gamma_0$ leads to an approximation of the form
$(\bm \gamma-\bm \gamma_0)'\bm M(\bm \gamma_0)(\bm \gamma-\bm
\gamma_0),$ where $i,j$ element of $\bm M(\bm \gamma)$ is given by
$\bm M(\bm
\gamma)_{(i,j)}=\sum_{l=1}^p\sum_{k=1}^p\frac{\partial}{\partial
  \theta_k \partial \theta_l} m(\bm \theta)\frac{\partial}{\partial
  \gamma_i}h_k(\bm \gamma)\frac{\partial}{\partial \gamma_j}h_l(\bm
\gamma)+ \sum_{k=1}^p\frac{\partial}{\partial \theta_k}m(\bm
\theta)\frac{\partial}{\partial \gamma_i \partial \gamma_j}h_k(\bm
\gamma)$, where $m(\bm \theta)=d^2(\bm \theta,\bm \theta_0)$ and $\bm
\theta=h(\bm \gamma)$. When evaluating this expression in the expansion point
the second summand vanishes as the gradient is zero. Hence one obtains
$\bm M(\bm \gamma)=\bm H(\bm \gamma)^T\bm V(h(\bm \gamma))\bm H(\bm
\gamma)$, which results in the density $p(\bm \gamma) \propto
\sqrt{\det(\bm H(\bm \gamma))^T\bm V(h(\bm \gamma))\det(\bm H(\bm
  \gamma))}= \det(\bm H(\bm \gamma))\sqrt{\det(\bm V(h(\bm \gamma)))}.$ $\Box$

\end{appendix}

\bibliographystyle{biom}
\bibliography{/home/bjoern/Projekte/BibFile/bibl}

\begin{thebibliography}{}

\bibitem[\protect\citeauthoryear{Balasubramanian}{Balasubramanian}{1997}]{bala%
:1997}
Balasubramanian, V. (1997).
\newblock Statistical inference, {O}ccam's razor, and statistical mechanics on
  the space of probability distributions.
\newblock {\em Neural Computation} {\bf 9,} 349--369.

\bibitem[\protect\citeauthoryear{Bates and Watts}{Bates and
  Watts}{1988}]{bate:watts:1988}
Bates, D.~M. and Watts, D.~G. (1988).
\newblock {\em Nonlinear Regression Analysis and Applications}.
\newblock John Wiley and sons, New York.

\bibitem[\protect\citeauthoryear{Berger, Bernardo, and Sun}{Berger
  et~al.}{2009}]{berg:bern:sun:2009}
Berger, J.~O., Bernardo, J.~M., and Sun, D. (2009).
\newblock The formal definition of reference priors.
\newblock {\em Annals of Statistics} {\bf 37,} 905--938.

\bibitem[\protect\citeauthoryear{Bornkamp}{Bornkamp}{2011}]{born:2011}
Bornkamp, B. (2011).
\newblock Approximating probability densities by iterated {L}aplace
  approximations.
\newblock {\em Journal of Computational and Graphical Statistics} {\bf 00,}
  00--00.

\bibitem[\protect\citeauthoryear{Bornkamp and Ickstadt}{Bornkamp and
  Ickstadt}{2009}]{born:icks:2009b}
Bornkamp, B. and Ickstadt, K. (2009).
\newblock A note on {B}-splines for semiparametric elicitation.
\newblock {\em The American Statistician} {\bf 63,} 373--377.

\bibitem[\protect\citeauthoryear{Bornkamp, Pinheiro, and Bretz}{Bornkamp
  et~al.}{2010}]{DoseFinding}
Bornkamp, B., Pinheiro, J., and Bretz, F. (2010).
\newblock {\em DoseFinding: Planning and Analyzing Dose Finding experiments}.
\newblock {R} package version 0.4-1.

\bibitem[\protect\citeauthoryear{Daniels}{Daniels}{1999}]{dani:1999}
Daniels, M.~J. (1999).
\newblock A prior for the variance in hierarchical models.
\newblock {\em Canadian Journal of Statistics} {\bf 27,} 567--578.

\bibitem[\protect\citeauthoryear{Dembski}{Dembski}{1990}]{demb:1990}
Dembski, W.~A. (1990).
\newblock Uniform probability.
\newblock {\em Journal of Theoretical Probability} {\bf 3,} 611--626.

\bibitem[\protect\citeauthoryear{Dragalin, Bornkamp, Bretz, Miller,
  Padmanabhan, Patel, Perevozskaya, Pinheiro, and Smith}{Dragalin
  et~al.}{2010}]{drag:2010}
Dragalin, V., Bornkamp, B., Bretz, F., Miller, F., Padmanabhan, S.~K., Patel,
  N., Perevozskaya, I., Pinheiro, J., and Smith, J.~R. (2010).
\newblock A simulation study to compare new adaptive dose-ranging designs.
\newblock {\em Statistics in Biopharmaceutical Research} {\bf 2,} 487--512.

\bibitem[\protect\citeauthoryear{Dryden, Koloydenko, and Zhou}{Dryden
  et~al.}{2009}]{dryd:kolo:zhou:2009}
Dryden, I.~L., Koloydenko, A., and Zhou, D. (2009).
\newblock Non-{E}uclidean statistics for covariance matrices with applications
  to diffusion tensor imaging.
\newblock {\em Annals of Applied Statistics} {\bf 3,} 1102--1123.

\bibitem[\protect\citeauthoryear{Ghosal, Ghosh, and Ramamoorthi}{Ghosal
  et~al.}{1997}]{ghos:ghos:rama:1997}
Ghosal, S., Ghosh, J.~K., and Ramamoorthi, R.~V. (1997).
\newblock Non-informative priors via sieves and packing numbers.
\newblock In Panchapakesan, S. and Balakrishnan, N., editors, {\em Advances in
  Statistical Decision Theory and Applications}, pages 119--132.
  Birkh{\"a}user, Boston.

\bibitem[\protect\citeauthoryear{Ghosh, Delampady, and Samanta}{Ghosh
  et~al.}{2006}]{ghos:dela:sama:2006}
Ghosh, J.~K., Delampady, M., and Samanta, T. (2006).
\newblock {\em An Introduction to Bayesian Analysis: Theory and Methods}.
\newblock Springer, New York.

\bibitem[\protect\citeauthoryear{Jeffreys}{Jeffreys}{1961}]{jeff:1961}
Jeffreys, H. (1961).
\newblock {\em Theory of Probability}.
\newblock Oxford University Press.

\bibitem[\protect\citeauthoryear{Jones, Plank, and Sleeman}{Jones
  et~al.}{2010}]{jone:plan:slee:2010}
Jones, D.~S., Plank, M.~J., and Sleeman, B.~D. (2010).
\newblock {\em Differential Equations and Mathematical Biology}.
\newblock Chapman and Hall, Boca Raton.

\bibitem[\protect\citeauthoryear{Kass and Wasserman}{Kass and
  Wasserman}{1996}]{kass:wass:1996}
Kass, R.~E. and Wasserman, L. (1996).
\newblock The selection of prior distributions by formal rules.
\newblock {\em Journal of the American Statistical Association} {\bf 91,}
  1343--1370.

\bibitem[\protect\citeauthoryear{Kolmogorov and Tihomirov}{Kolmogorov and
  Tihomirov}{1961}]{kolm:tiho:1961}
Kolmogorov, A.~N. and Tihomirov, V.~M. (1961).
\newblock $\epsilon-$entropy and $\epsilon-$capacity of sets in function
  spaces.
\newblock {\em American Mathematics Society Translations Ser. 2} {\bf 17,}
  277--364.

\bibitem[\protect\citeauthoryear{Leydold and H\"ormann}{Leydold and
  H\"ormann}{2010}]{leyd:hoer:2010}
Leydold, J. and H\"ormann, W. (2010).
\newblock {\em Runuran: R interface to the UNU.RAN random variate generators}.
\newblock {R} package version 0.15.0.

\bibitem[\protect\citeauthoryear{Lindsey}{Lindsey}{2001}]{lind:2001}
Lindsey, J.~K. (2001).
\newblock {\em Nonlinear Models for Medical Statistics}.
\newblock Oxford University Press, Oxford.

\bibitem[\protect\citeauthoryear{M{\"u}ller, Berry, Grieve, and
  Krams}{M{\"u}ller et~al.}{2006}]{muel:berr:grie:2006}
M{\"u}ller, P., Berry, D.~A., Grieve, A.~P., and Krams, M. (2006).
\newblock A {B}ayesian decision-theoretic dose-finding trial.
\newblock {\em Decision Analysis} {\bf 3,} 197--207.

\bibitem[\protect\citeauthoryear{Neuenschwander, Capkun-Niggli, Branson, and
  Spiegelhalter}{Neuenschwander et~al.}{2010}]{neue:capk:bran:2010}
Neuenschwander, B., Capkun-Niggli, G., Branson, M., and Spiegelhalter, D.~J.
  (2010).
\newblock Summarizing historical information on controls in clinical trials.
\newblock {\em Clinical Trials} {\bf 7,} 5--18.

\bibitem[\protect\citeauthoryear{O'Hagan, Buck, Daneshkhah, Eiser, Garthwaite,
  Jenkinson, Oakley, and Rakow}{O'Hagan et~al.}{2006}]{ohag:2006}
O'Hagan, A., Buck, C., Daneshkhah, A., Eiser, R., Garthwaite, P., Jenkinson,
  D., Oakley, J., and Rakow, T. (2006).
\newblock {\em Uncertain Judgements: Eliciting Expert Probabilities}.
\newblock John Wiley and Sons Inc.

\bibitem[\protect\citeauthoryear{O'Hagan and Forster}{O'Hagan and
  Forster}{2004}]{ohag:fors:2004}
O'Hagan, A. and Forster, J. (2004).
\newblock {\em Kendall's Advanced Theory of Statistics, Volume 2B: {B}ayesian
  Inference}.
\newblock Arnold, London, 2nd edition.

\bibitem[\protect\citeauthoryear{Paulo}{Paulo}{2005}]{paul:2005}
Paulo, R. (2005).
\newblock Default priors for {G}aussian processes.
\newblock {\em Annals of Statistics} {\bf 33,} 556--582.

\bibitem[\protect\citeauthoryear{Pennec}{Pennec}{2006}]{penn:2006}
Pennec, X. (2006).
\newblock Intrinsic statistics on {R}iemannian manifolds: Basic tools for
  geometric measurements.
\newblock {\em Journal of Mathematical Imaging and Vision} {\bf 25,} 127--154.

\bibitem[\protect\citeauthoryear{{R Development Core Team}}{{R Development Core
  Team}}{2011}]{R}
{R Development Core Team} (2011).
\newblock {\em R: A Language and Environment for Statistical Computing}.
\newblock R Foundation for Statistical Computing, Vienna, Austria.
\newblock {ISBN} 3-900051-07-0.

\end{thebibliography}

\end{document}